\documentclass[superscriptaddress,aps,prb,noeprint,reprint,floatfix]{revtex4-2}
\usepackage{graphicx}
\usepackage{tabularx}
\usepackage{amssymb}
\usepackage{amsmath}
\usepackage{dcolumn}
\usepackage[utf8]{inputenc}
\usepackage{bm}
\usepackage{tikz}
\usepackage{color}
\usepackage{colortbl}
\usepackage{xcolor}
\usepackage{float}

\definecolor{Az75}{RGB}{4,222,255}
\definecolor{Trq67}{RGB}{2,208,116}
\definecolor{r1}{RGB}{255,0,114}
\definecolor{Pnk102}{RGB}{255,0,166}
\definecolor{Blu82}{RGB}{134,134,255}

\begin{document}
\title{The role of density-dependent magnon hopping and magnon-magnon repulsion   
in ferrimagnetic spin-(1/2, $S$) chains in a magnetic field}
\author{W. M. da Silva}
\affiliation{Secretaria de Educa\c{c}\~ao da Para\'{\i}ba, 58015-900 Jo\~ao Pessoa-PB, Brasil}
\affiliation{Laborat\'{o}rio de F\'{i}sica Te\'{o}rica e Computacional, Departamento de F\'{i}sica, Universidade Federal de Pernambuco, 50760-901 Recife-PE, Brasil}
\author{R. R. Montenegro-Filho}
\affiliation{Laborat\'{o}rio de F\'{i}sica Te\'{o}rica e Computacional, Departamento de F\'{i}sica, Universidade Federal de Pernambuco, 50760-901 Recife-PE, Brasil}

\begin{abstract}
We compare the ground-state features of alternating ferrimagnetic 
chains $(1/2, S)$ with $S=1,3/2,2,5/2$ in a magnetic field and the corresponding Holstein-Primakoff bosonic models 
up to order $\sqrt{s/S}$, with $s=1/2$, considering the fully polarized magnetization as the boson vacuum. {The single-particle Hamiltonian is a Rice-Mele model with 
uniform hopping and modified boundaries, while the interactions have a correlated (density-dependent) hopping term and magnon-magnon repulsion.} The magnon-magnon repulsion increases the many-magnon energy and the density-dependent hopping decreases the kinetic energy. We use density matrix renormalization group calculations to investigate the effects of these two interaction terms in the bosonic model{, and display the quantitative agreement between the results from the spin model and the full bosonic approximation.  In particular, we verify the good accordance in the behavior of the edge states, associated with the ferrimagnetic plateau, from the spin and from the bosonic models. Furthermore, we show that the boundary magnon density strongly depends on the interactions and particle statistics.}.\end{abstract}

\maketitle

\section{Introduction}

The gapped phases of magnetic insulators are responsible for magnetization ($m$) plateaus
in the $m$ vs. magnetic field curves \cite{Giamarchi2008}. In one dimension, these incompressible phases satisfy the
topological Oshikawa-Yamanaka-Affleck \cite{PhysRevLett.78.1984} criteria, and exhibit associated edge states in open spin chains 
\cite{Hu2014,*Hu2015}. The gapped phases are separated by gapless phases that have a 
low-energy physics described by the Tomonoga-Luttinger-liquid theory \cite{giamarchi2003quantum,*PhysRevB.55.5816}. 
Thus, the magnetic field $h$ induces quantum phase transitions in the spin chain, with quantum critical points
at the plateau extremes \cite{sachdev2001quantum,Vojta2003}. In the vicinity of the quantum critical points, the magnons are in
a high-dilute regime and can be treated as a gas of hard-core bosons \cite{Affleck91,Sorensen1993} or fermions \cite{Tsvelik90}. 
In this approximation, the energy is comprised only by a \textit{simple hopping term} 
and the uniform Zeeman term, which plays the role of chemical potential in the effective model.
Following this approach, we can show that the $m(h)$ curve (magnon density) presents a square-root singularity in the gapless side 
of the transition. The first correction to this law is linear and obtained by taking into account \textit{magnon-magnon interaction} 
through a phenomenological scattering length \cite{Okunishi1999,Lou2000,Affleck2004,Affleck2005,Vanderstraeten2015}.  

Another kind of hopping term is known as \textit{correlated} (or \textit{density-dependent})
\textit{hopping} and is essential in the modeling of a variety of quantum systems \cite{Dobry2011}.
One of these terms, the bond-charge interaction \cite{Kivelson1987}, is used to model
electrons in strongly correlated materials \cite{Hirsch1989,HirschPRB1989,Japaridze1999,Vidal2001,Anfossi2007} and was 
particularly investigated in the context of high-$T_c$ superconductivity. Besides, the 
Hubbard model with this term has an exact solution in a special point of the parameter
space \cite{Arrachea1994,Schadschneider1995,Dolcini2002,Vitoriano2009}. 
On the other hand, the extended boson Hubbard model with a density-dependent hopping is an effective Hamiltonian for bosonic molecules, typically polar species \cite{Maik2013,Baier2016,Fazzini2017}, in 
optical lattices \cite{Bendjama2005,Schmidt2006,Schmidt2008,Eckholt2009,Jiang2012,Jurgensen2014,Biedron2018,Johnstone2019}.
Further, general correlated hopping hard-core bosonic Hamiltonians are investigated to understand the physics 
of frustrated insulating magnetic materials \cite{Bendjama2005,Schmidt2006,Schmidt2008,Stasinska2019}. 

The ground-state of ferrimagnetic chains satisfy the Lieb-Mattis theorem \cite{Lieb.Mattis} 
{ and exhibit ferromagnetic and antiferromagnetic long-range orders \cite{Tian}}. and was investigated in the
isotropic \cite{PhysRevLett.78.4853,*PhysRevB.59.14384,AlcarazandMa,Gu2006,*PhysRevB.80.014413,*Gong2010} and
anisotropic \cite{AlcarazandMa,YamamotoPRB99,Verissimo2019} cases. 
Ferrimagnetic systems  \.
In particular, the behavior of 
the edge states associated with the 1/3 magnetization plateau of the AB$_2$ anisotropic 
chain was recently investigated \cite{Montenegro-Filho2020}.
Furthermore, rich phase diagrams are 
observed through doping \cite{PhysRevLett.74.1851,RenePRB2006,PhysRevB.59.7973,
Montenegro-Filho2014,Kobayashi2016} 
or adding geometric frustration 
\cite{Hida1994,Takano1996,RenePRB2008,Ivanovart10,Shimokawa2011,Furuya2014,
Amiri2015,Hida2017}
to the ferrimagnetic models. In particular,  
ferrimagnetic spin-(1/2, $S$) chains under an applied magnetic field present 
magnetization plateaus at $m=S-1/2$ (ferrimagnetic plateau) and $m=S+1/2$ (saturation plateau),
where $m$ is the magnetization per unit cell 
\cite{PatiJPCM1997,*PhysRevB.55.8894,Maisinger1998,ReneJPCM2011,Strecka2017,StreckaActa2017,DaSilva2017}.
On the experimental side, the one-dimensional magnetic phase of a variety of bimetallic compounds was shown to be modeled
by spin-(1/2, $S$) ferrimagnetic chains \cite{JPSJ.67.2209,JPSJ.68.2214,Kahn1988,Gleizes1981,Verdaguer1984,VanKoningsbruggen1990}.
Recently, the full magnetization curve of the charge-transfer salt 
(4-Br-$o$-MePy-V)FeCl$_4$ was experimentally investigated \cite{Yamaguchi2020,*YamaguchiPRB2020} and shown to be 
a spin-(1/2, 5/2) chain above the three-dimensional ordering temperature.  

Since spin-($s$, $S$) ferrimagnetic chains have a long-ranged ordered ground state, 
linear and interacting spin-wave theory \cite{Noriki2017} from the classical \textit{ferrimagnetic state} 
was used to characterize their low-energy magnetic excitations, mainly through the
Holstein-Primakoff formalism \cite{PatiJPCM1997,*PhysRevB.55.8894,Brehmer1997,PhysRevB.57.R14008,PhysRevB.57.13610,Maisinger1998,JCPM.10.11033.1998,Ivanov2000,PhysRevB.69.06}. Furthermore, linear spin-wave theory from the \textit{fully polarized state} \cite{DaSilva2017}  
gives good results for the gapped and gapless phases of spin-(1/2, $S$) chains in a magnetic field. 

Here we show that the Holstein-Primakoff Hamiltonian up to order $\sqrt{s/S}$ gives almost exactly results 
to the ground-state phase diagram of ferrimagnetic spin-(1/2, S) chains in a magnetic field. 
We use the density matrix renormalization group (DMRG) \cite{WhitePRB93,WhitePRL1992} to obtain 
the magnetization curves, besides local properties, from the Holstein-Primakoff Hamiltonian 
and the spin Hamiltonian. In addition to the \textit{simple hopping} and 
the \textit{nearest-neighbor} interacting terms, a \textit{correlated hopping} term is essential to obtain a 
good accord between the numerical results from the spin and bosonic models. In Sec. \ref{sec:hp} we present 
the Holstein-Primakoff Hamiltonian and the analytical formulas for the hard-core boson approximation.  
In Sec. III we compare the DMRG results for the magnetization and local properties from the Holstein-Primakoff Hamiltonian and 
the spin model. In Sec. IV we present a summary of the main results of the manuscript.  

\section{Holstein-Primakoff bosonic Hamiltonian from the fully polarized vacuum}
\label{sec:hp}

\begin{figure}[htb]
\includegraphics*[width=0.4\textwidth]{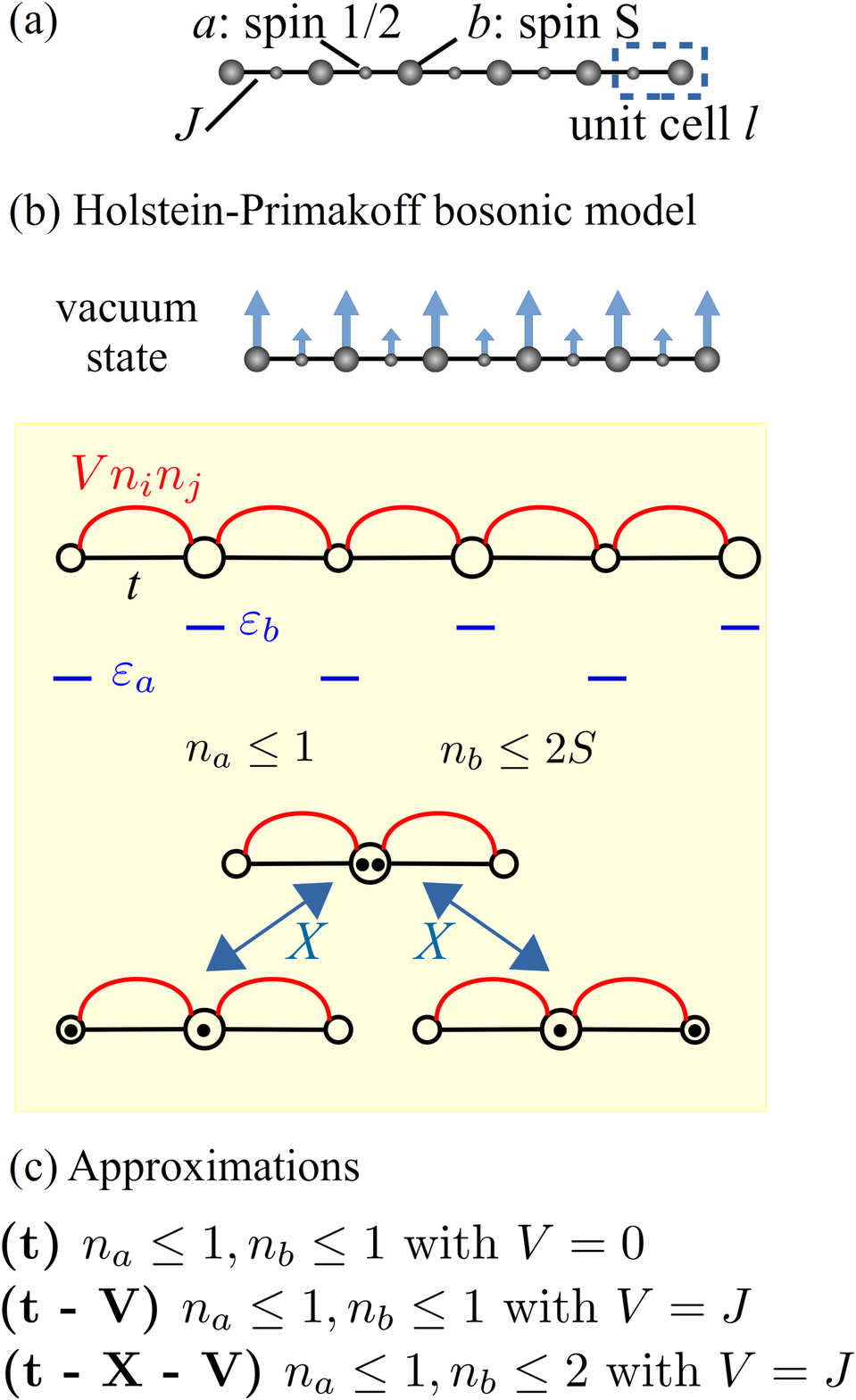}
\caption{(a) Schematic representation of the alternating spin model, with spin-$1/2$ at $a$ sites, and spin-$S$ 
at $b$ sites, with $S=1,~3/2,~2,\text{ and } 5/2$. The $z$-direction is the direction of an applied magnetic 
field $B$ and the superexchange coupling is $J$. (b) Holstein-Primakoff bosonic Hamiltonian 
up to $\mathcal{O}(S^{-1/2})$, with a hopping term $t=J\sqrt{sS}$, local potentials 
$\varepsilon_a=-2SJ$ and $\varepsilon_b=-2sJ$, nearest-neighbor interaction $V=J$, 
and density dependent correlated hopping process $X=J\sqrt{s/S}$. The $a$ sites have a hard-core constraint, 
$n_a\leq 1$, while the constraint $n_b\leq 2S$ is imposed on $b$ sites, with $n_a$ ($n_b$) as the 
number of bosons in one $a$ ($b$) site. The magnetic field $B$ acts as a chemical potential $\mu$ in the 
bosonic model: $\mu=-B$. (c) The bosonic approximations investigated: $t$, $t-V$, and $t-X-V$. In the 
$t$ and $t-V$ approximations, there is a hard-core constraint on all sites; while in the $t-X-V$ model, 
the $b$ sites can accommodate up to two magnons.}
\label{fig:models}
\end{figure}

The alternating mixed-spin $(s=1/2,S)$ chain with $L$ unit cells has the Hamiltonian 
\begin{equation}
 \mathcal{H}^{\textsc{spin}}=J\sum_{j=1}^L\left(\mathbf{s}_{j}\cdot\mathbf{S}_j+\mathbf{s}_{j}\cdot\mathbf{S}_{j-1}\right)-BS^z_{tot},
\label{eq:hspin}
 \end{equation}
where 
\begin{equation}
S^z_{tot}=\sum_j(s^z_j+S^z_j) 
\end{equation}
is the $z$-component of the total spin, 
and we consider the magnetic field $B$ in the $z$ direction, with $g\mu_B\equiv 1$, 
where $\mu_B$ is the Bohr magneton and $g$ is the $g$-factor. 
The spin$-1/2$ are attached to $a$ sites, while spin$-S$ to 
$b$ sites along the chain, and we study chains for which $S=1,~3/2,~2,~5/2$, 
as schematically shown in Fig. \ref{fig:models}(a). The ground-state total spin 
for $B=0$ is $S-s$ and has a copy in each sector in the range
$-(S-s) \leq S^z_{tot}\leq (S-s)$, as expected from the Lieb-Mattis theorem
\cite{Lieb.Mattis}, with a ferrimagnetic (FRI) long-range ordered state.
If a little 
magnetic field is applied, the ground state with $S^z_{tot}=(S-s)$ is chosen. 
Further, the ground state has a finite gap to spin excitations 
carrying a spin $\Delta S^z=+1$, which induces a magnetization plateau at  
$m_{FRI}=(S-s)$. Also, a second magnetization plateau is the fully polarized 
plateau at $m_{FP}=S+s$. 

The \textit{fully polarized state} is an exact ground state of the spin Hamiltonian, 
and we build the spin-wave theory considering it as the magnon vacuum. 
Making the Holstein-Primakoff mapping on $a$ sites
\begin{eqnarray}
  s_{j}^z & = & s - n^a_j,\nonumber \\
  s_{j}^{-}&=&\sqrt{2s}a_j^\dag\Big(1-\frac{n^a_j}{2s}\Big)^{1/2}\approx \sqrt{2s}a_j^\dag\Big(1-\frac{n^a_j}{4s}\Big),
 \end{eqnarray}
and on $b$ sites:
\begin{eqnarray}
  S_{j}^z & = & S - n^b_j;\nonumber \\
  S_{j}^{-}&=&\sqrt{2S}b_j^\dag \Big(1-\frac{n^b_j}{2S}\Big)^{1/2}\approx \sqrt{2S}b_j^\dag \Big(1-\frac{n^b_j}{4S}\Big),
\end{eqnarray}
where $n^a_j=a^\dag_j a_j$ and $n^b_j=b^\dag_j b_j$, we arrive in the following spin-wave Hamiltonian  
 
\begin{widetext}
\begin{eqnarray}
 \mathcal{H}^{\text{\textsc{SW}}} & = & J\sum_{j}\Bigg\{\Big(S - n^b_j\Big)\Big(s - n^a_j\Big) + \sqrt{sS}\Bigg[\Big(1 -\frac{n^b_j}{4S}\Big)b_ja_j^\dag\Big(1 -\frac{n^a_j}{4s}\Big) +\nonumber \\
             & + & b_{j}^\dag\Big(1 -\frac{n^b_j}{4S}\Big)\Big(1 -\frac{n^a_j}{4s}\Big)a_{j}\Bigg] + \sqrt{sS}\Bigg[\Big(1 -\frac{n^a_j}{4s}\Big)a_j b_{j+1}^\dag\Big(1 -\frac{n^b_{j+1}}{4S}\Big) +\nonumber  \\ 
             & + & a_j^\dag\Big(1 -\frac{n^a_j}{4s}\Big)\Big(1 -\frac{n^b_{j+1}}{4S}\Big)b_{j+1}\Bigg] + \Big(s - n^a_j\Big)\Big(S - n^b_{j+1}\Big)\Bigg\} +\nonumber  \\
             & - & B\sum_j\Big(S + s - n^a_j - n^b_j\Big)+ \mathcal{O}(S^{-1}).
 \label{eq:HamiltMagnons_FP}
\end{eqnarray}
\end{widetext}
Dropping the classical energy of the ferromagnetic state $E_{class} = 2JLsS -B\big(S+s\big)L$, the relevant 
magnon Hamiltonian is:
\begin{equation}
 {H}^{\text{\textsc{SW}}} = H_t + H_X + H_V+\mathcal{O}(S^{-1}),
 \label{eq:hamiltonianhp}
\end{equation}
with $H^{\text{\textsc{SW}}}= \mathcal{H}^{\text{\textsc{SW}}}-E_{class}$.

As sketched in Fig. \ref{fig:models}(b), the $H_t$ term comprises a hopping process and distinct local potentials on
$a$ and $b$ sites: 
\begin{eqnarray}
H_t &=& t\sum_{j}\Big(b_{j}^\dag a_{j+1} + b_{j+1}^\dag a_{j+1}+\text{H. c.}\Big)\nonumber\\
& &+\sum_j[(\varepsilon_a-\mu) n^a_j + (\varepsilon_b-\mu)n^b_j] 
\label{eq:hamiltonianht}
\end{eqnarray}
with 
\begin{equation}
 \begin{cases}
  t=J\sqrt{sS}\text{, hopping parameter;}\\
  \varepsilon_a=-2SJ\text{, local potential on $a$ sites;}\\
  \varepsilon_b=-2sJ\text{, local potential on $b$ sites;}\\
  \mu=-B.
 \end{cases}
 \label{eq:httransform}
\end{equation}
In an open chain, if $a$ or $b$ is a boundary site, the local potential  
is half of the above value: 
\begin{equation}
 \begin{cases}
  \varepsilon^{\text{(boundary)}}_a=-SJ\\
  \varepsilon^{\text{(boundary)}}_b=-sJ.
 \end{cases}
 \label{eq:localboundary}
\end{equation}
Considering the local potentials, we see that the magnon has a higher probability to be found on $a$ sites, and this probability increases with $S$. However, since the hopping parameter $t$ increases with 
$\sqrt{S}$, quantum fluctuations are relevant for moderate values of $S$, and the magnons can 
overcome the potential barrier between $a$ and $b$ sites.

{ We observe that the bulk Hamiltonian $H_t$ is a particular case 
of the Rice-Mele model \cite{Rice1982}:
\begin{eqnarray}
H^{\text{\scriptsize Rice-Mele}}&=&\sum_{j}\Big(t_2 b_{j}^\dag a_{j+1} + t_1 b_{j+1}^\dag a_{j+1}+\text{H. c.}\Big)\nonumber\\&+&\sum_j(\varepsilon_a n^a_j + \varepsilon_b n^b_j),
\label{eq:ricemele}
\end{eqnarray}
putting $\mu=0$ and considering the general case of alternating hopping: $t_1$ ($t_2$) for
intra-cell (inter-cell) hopping. The Rice-Mele Hamiltonian was originally proposed 
to model the physics of electrons in polymers, but  
is a paradigmatic model to the understanding of topological insulators,
and can be realized by atoms in optical lattices \cite{Cooper2019}. The model presents the bulk-boundary 
correspondence \cite{Lin2020}, and an interacting version was recently investigated \cite{Lin2020} to  probe the connection between topology and particle-particle interactions \cite{Lin2020}. The Rice-Mele model recovers the bulk Hamiltonian $H_t$ for $t_1=t_2$.
However, for an open chain, the mapping of the spin model requires local potentials on
the boundary sites, Eq. (\ref{eq:localboundary}).}

The second term in the bosonic Hamiltonian (\ref{eq:hamiltonianhp}) 
is a density-dependent or correlated hopping term given by
\begin{equation}
H_X  = -X \sum_j\big[(a^\dag_j+a^\dag_{j+1})n^b_jb_j+\text{H. c.}\big],
\label{eq:hamiltonianx}
\end{equation}
with 
\begin{equation}
X=\frac{J}{4}\sqrt{\frac{s}{S}}.
\label{eq:xterm}
\end{equation}
Since $s=1/2$ and $n^a_j\leq1$, a hard-core constraint must be imposed on
$a$ sites for all models considered. Hence, we discard a term similar to the $X$ term but with $a$ and $b$ variables
exchanged, with $X^\prime=\frac{J}{4}\sqrt{S/s}$ as the correlated hopping amplitude.
As sketched in Fig. \ref{fig:models}(b),  
the energy of the system is lowered by the hopping of a magnon to a site that 
is already occupied. In other words, the magnon probability to overcome the 
potential barrier between $a$ and $b$ sites increases if there is one magnon on the $b$ site. 
This term becomes relevant for 
higher magnon densities, since it is an interaction term, and $X\rightarrow 0$ as $S$ increases.

The last term in the Hamiltonian (\ref{eq:hamiltonianhp}), see Fig. \ref{fig:models}(b), is a repulsive term between magnons in nearest neighbor sites:
\begin{equation}
H_V=V\sum_j n^a_{j+1}(n^b_j + n^b_{j + 1}),
\label{eq:hamiltonianhv}
\end{equation}
with $V=J$, and increases the energy for higher magnon densities. 
For one magnon per unit cell, this term favors the magnon localization on 
alternating $a$ sites, since the local potentials ($\varepsilon$) favors
the occupation of $a$ sites.
Thus, $\varepsilon$ (any density) and $V$ (higher densities) favor 
magnon localization on $a$ sites, while quantum fluctuations 
(tunneling between $a$ and $b$ sites) are favored by 
$t$ (any density) and $X$ (higher densities).

In this work, we compare data from three approximations of Hamiltonian 
(\ref{eq:hamiltonianhp}), as summarized in Fig. \ref{fig:models}(c), and 
from the spin Hamiltonian (\ref{eq:hspin}).  The first, which we 
identify as the $t$ approximation, 
is an analytical solution 
to the free hard-core model. In this approximation, all 
sites have a hard-core constraint and can be occupied by only one magnon: 
$n^a_j\leq 1$ and $n^b_j\leq 1$, for any $j$. This constraint implies that there is not 
an energy contribution from the $X$ term, and we drop the nearest-neighbor interaction $V$.
In the second approximation, $t-V$, we keep the hard-core constraint but 
add the $V$ contribution to $H_t$. The last approximation, $t-X-V$, has 
the three terms $H_t$, $H_X$, and $H_V$. We consider a hard-core constraint on $a$ sites
and a constraint $n^b_j\leq 2$ on $b$ sites. As we present below, the relaxation
of the hard-core constraint on $b$ sites and the consequent activation of the correlated hopping 
term implies an excellent agreement between the results of this approximation 
and the spin model.

\subsection{Free hard-core magnons [$t$-approximation]: $H_t$ and {$L\rightarrow\infty$}}

The first approximation to the magnetization curve, a many-magnon state,
is to consider the magnons as free hard-core bosons or free fermions. This 
mapping is exact in the high-dilute regime of magnons, near the  
saturation field. Here, we extend this approach for the full range of 
the magnetization, $(S-s) \leq m \leq (S+s)$, and compare 
their results to more precise calculations considering the interaction terms. 

The single-magnon energies are given by the term $H_t$, Eq. (\ref{eq:hamiltonianht}), in Hamiltonian 
(\ref{eq:hamiltonianhp}).
Using the following Fourier transforms:
\begin{equation}
\begin{cases}
a_{j}= \frac{1}{\sqrt{L}}\sum_{k}e^{-ik/4}e^{ikj}a_{k};\\
b_{j}= \frac{1}{\sqrt{L}}\sum_{k}e^{+ik/4}e^{ikj}b_{k},
\end{cases}
\label{eq:Fourier}
\end{equation}
where a phase $e^{\pm ik/4}$ is included to ease the calculation,  $H_t$ becomes:
\begin{equation}
 H_t=\sum_k \begin{pmatrix} a^\dagger_k & b^\dagger_k \end{pmatrix} 
\begin{pmatrix} \varepsilon_a & \gamma_k \\ \gamma_k & \varepsilon_b \end{pmatrix}
\begin{pmatrix} a_k \\ b_k \end{pmatrix},
\end{equation}
where 
\begin{equation}
\gamma_k=2t\cos(k/2). 
\end{equation} 
After diagonalization, the Hamiltonian is written as 
\begin{equation}
H_t = \sum_{k}\omega^{\scriptscriptstyle{(-)}}_{k}n^{\scriptscriptstyle{(-)}}_{k}+\omega^{\scriptscriptstyle{(+)}}_k n^{\scriptscriptstyle{(+)}}_{k},
\label{eq:htdiagonal}
\end{equation}
where the dispersion relations are
\begin{equation}
 \omega^{\scriptscriptstyle{(\pm)}}_k=\frac{\varepsilon_a+\varepsilon_b}{2}\pm\omega_k=-J(s+S)\pm\omega_k,
 \label{eq:dispersion}
\end{equation}
with 
\begin{align}
 \omega_k&=\sqrt{\left(\frac{\varepsilon_b-\varepsilon_a}{2}\right)^2+\gamma_k^2}\nonumber\\
         &=J\sqrt{(S-s)^2+4sS\cos^2(k/2)},
\end{align}
and $n^{\scriptscriptstyle{(\pm)}}_k=\alpha^{\scriptscriptstyle{(\pm)}\dag}_k\alpha^{\scriptscriptstyle{(\pm)}}_k$, where 
\begin{equation}
\begin{pmatrix}\alpha^{\scriptscriptstyle{(-)}}_k \\ \alpha^{\scriptscriptstyle{(+)}}_k\end{pmatrix}=
\begin{pmatrix}\cos\theta_k & -\sin\theta_k\\ \sin\theta_k & \cos\theta_k\end{pmatrix}
\begin{pmatrix}a_k \\ b_k\end{pmatrix},
\label{eq:transformation}
\end{equation}
with
\begin{equation}
\begin{cases}
 \cos^2\theta_k = \frac{1}{2}+\frac{S-s}{{2\omega}_k};\\
 \sin^2\theta_k = \frac{1}{2}-\frac{S-s}{{2\omega}_k}.
\end{cases}
\label{eq:solution}
\end{equation}
A magnetic field (or chemical potential) adds a $+B$ energy term to the two bands.
\begin{figure}[htb]
\includegraphics*[width=0.48\textwidth]{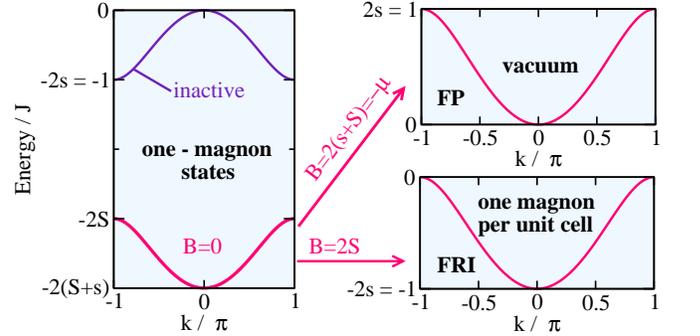}
\caption{One-magnon states from the fully polarized (FP) vacuum
and $B=0$ as a function of lattice wave-vector $k$. In the $t$-approximation,  
the high energy band is inactive. For $B=2(s+S)=-\mu$, the exact ground
state is the FP vacuum; while there is one magnon per unit cell in the system for $B=2S$,
and the ground state is ferrimagnetic (FRI).}
\label{fig:magnonbands}
\end{figure}

The magnon bands are shown in Fig. \ref{fig:magnonbands} for (i) $B=0$; (ii) 
$B=2(s+S)$, the saturation 
field, for which $m=m_{\text{\scriptsize{FP}}}=(S+s)$; and (iii) $B=2S$, 
the critical field of the ferrimagnetic
plateau, for which $m=m_{\text{\scriptsize{FRI}}}=(S-s)$, 
or 1 magnon per unit cell: 
$m_{\text{\scriptsize{FP}}}-m_{\text{\scriptsize{FRI}}}=(S+1/2)-(S-1/2)=1$. 
As expected from the Lieb-Mattis theorem, the magnetization
in the null field is $S-s$. 
In the free hard-core boson or free fermion approximation, we fill the single-particle 
states following a Fermi distribution up to the effective Fermi wave-vector $k_F$.
As shown in Fig. \ref{fig:magnonbands}, if we follow this procedure 
the two bands should be filled for $B=0$ (two magnons per unit cell) 
and the magnetization curves of the spin model 
would not be reproduced. We have shown in Ref. [\onlinecite{DaSilva2017}] that this 
problem can be overcome, even for finite $T$, by 
introducing an effective chemical potential $\mu$ to the upper band, in a way similar to 
Takahashi's solution to the ferromagnetic linear chain \cite{PhysRevLett.58.168}.
In particular, $\mu\rightarrow -J=-2sJ$ as $T\rightarrow 0$, such that the overall effect of  
$\mu$ at $T=0$ is the suppression of the upper band. 
Hence, in the free hard-core approximation, we must consider only 
the lower band in the calculations, as schematically indicated 
in Fig. \ref{fig:magnonbands}.
Thus, for example, the energy per unit cell in the free hard-core 
approximation is written as
\begin{equation}
\frac{E}{L}=\frac{1}{L}\sum_{k=-k_F}^{k_F}\omega^{\scriptscriptstyle{(-)}}_{k}
\label{eq:maghc}
\end{equation}
for a magnon density per unit cell $n$, where
\begin{equation}
 k_F=\pi n,
\label{eq:kf}
\end{equation}
and $n=m_{\text{\scriptsize{FP}}}-m$.

\subsubsection{Average local magnetizations}

If the chain has a magnon density per unit cell $n$ and considering the hard-core 
approximation, the average magnetizations of $a$ and $b$ sites are given by 
\begin{equation}
\begin{cases}
   \langle s^z \rangle = s - \frac{1}{L}\sum_{k=-k_F}^{k_F} \langle n^a_k \rangle&\text{($a$ sites)};\\
   \langle S^z \rangle = S - \frac{1}{L}\sum_{k=-k_F}^{k_F} \langle n^b_k\rangle&\text{($b$ sites)}.
\end{cases}
\label{eq:magnetization}
\end{equation} 
Using Eqs. (\ref{eq:transformation}) and discarding terms involving the upper band, 
we obtain
\begin{equation}
\begin{cases}
\langle n^a_k \rangle = n_k^{\scriptscriptstyle{(-)}}\cos^2\theta_k;\\
\langle n^b_k \rangle = n_k^{\scriptscriptstyle{(-)}}\sin^2\theta_k.
\end{cases}
\label{eq:nsnk} 
\end{equation}
We, thus, have
\begin{equation}
\begin{cases}
\big \langle s^z \big \rangle = s - \frac{1}{2} - \frac{S - s}{2}\sum_{k=-k_F}^{k_F} \frac{1}{\omega_{k}};\\
\big \langle S^z \big \rangle = S - \frac{1}{2} + \frac{S - s}{2}\sum_{k=-k_F}^{k_F} \frac{1}{\omega_{k}},
\end{cases}
\label{eq:localmag}
\end{equation}
after the substitution of Eqs. (\ref{eq:solution}) in Eqs. (\ref{eq:nsnk}) and the results 
in Eqs. (\ref{eq:magnetization}).

\section{DMRG results for the spin model and the bosonic Hamiltonians}

\subsection{Methodology}
We use the density matrix renormalization group to obtain the magnetization curves 
and local properties of  
the spin, $t-V$, and $t-X-V$ models; 
and also compare this 
data with the analytical results from the free hard-core model ($t$-model) 
in the thermodynamic limit: $L\rightarrow\infty$.  
These approximations are summarized in Fig. \ref{fig:models}(c). All DMRG results (boson and spin models) 
are obtained through the Algorithms and Libraries for Physics Simulations (ALPS) project 
\cite{Bauer2011} for chains with $L=128$ unit cells, with one $a$ site at one extreme and a $b$ 
site at the other. 
{ If the system has an $a$ site at the left extreme and a $b$ in the right ($a$-$b$ boundaries), the renormalization process for the magnetization step inside the ferrimagnetic plateau becomes trapped in a metastable state for $S=3/2,2,\text{ and }5/2$. In these cases, the global energy minimum is reached by the algorithm if the chains have a $b$ site at the left extreme and an $a$ site at the right extreme ($b$-$a$ boundaries). In the other magnetizations, this is irrelevant, i. e., the same state is calculated for the $a$-$b$ or the $b$-$a$ boundaries.}
We retain a maximum of 243 states per block
and the maximum discarded weight less than $10^{-9}$.

For the spin model, the magnetization curves are calculated 
from the lowest energy state for a fixed $S^z$ and $B=0$: $E(S^z)$; 
since for $B\neq0$ we need only to add the Zeeman term, such that 
$E_{B}(S^z)=E(S^z)-BS^z$.
In a gapless (non-plateau) phase, the magnetization curves are
made of finite steps in a finite-size system. Defining the extreme points of these 
finite steps as $B_-$ and $B_+$, we have 
\begin{equation}
B_{\pm}=\pm[E(S^z\pm1)-E(S^z)]
\end{equation} 
for the step at $S^z$. In a gapless phase, $B_-\rightarrow B_+$ as $L\rightarrow \infty$, while in a thermodynamic-limit 
plateau state $B_-\neq B_+$ for $L\rightarrow \infty$. In the last case, $B_{\pm}$ are quantum critical 
fields separating the plateau insulating state from a gapless critical Luttinger liquid phase. 
For the bosonic models, the magnon density 
per unit cell $n$ as a function of chemical potential $\mu$  
is obtained with a similar procedure. We calculate the lowest energy state for 
a fixed number of bosons $N$, with $N=nL$ and $\mu=0$. The value of the chemical 
potential at the extremes ($\mu_-$ and $\mu_+$) of the step at $N$ are given by  
\begin{equation}
\mu_{\pm}=\pm[E(N\pm1)-E(N)]. 
\end{equation}
A gapless phase has $\mu_+\rightarrow\mu_-$ as $L\rightarrow\infty$, while in a 
plateau insulating phase
$\mu_+\neq\mu_-$ in the thermodynamic limit. The transformation from the boson to the spin language is 
performed through the following equations: 
\begin{equation}
 \begin{cases}
  n=m-m_{\text{\scriptsize{FP}}},\text{ and}\\
  B=-\mu.
 \end{cases}
\end{equation}

\subsection{Magnetization curves and local magnetizations}
\begin{figure}
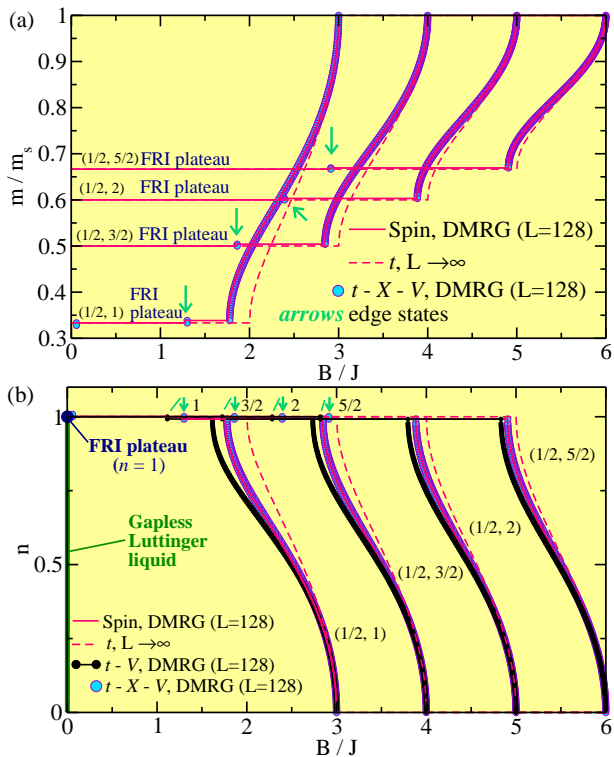

\includegraphics*[width=0.45\textwidth]{fig3a.eps}
\includegraphics*[width=0.45\textwidth]{fig3b.eps}
\caption{(a) Magnetization per unit cell ($m$) of $(1/2,S)$ chains, with $S=1,3/2,2,\text{ and }5/2$, normalized by
its saturation value ($m_s$) and (b) magnon densities per unit cell $n$ as functions of $B$ {in units of $J$}. 
DMRG data for (full lines) the spin model   
and for ($\color{Az75} \bullet$) the $t-X-V$ approximation.
We present in both figures the results for the $t$-model (dashed lines) 
and $L\rightarrow\infty$. 
In (b), we also show the magnon density as a function of $B$ {units of $J$} 
for ($\bullet$) the $t-V$ approximation, 
calculated with DMRG for a system with $L=128$. In the thermodynamic limit, 
the ferrimagnetic plateau is observed for $n=1$, while the 
gapless Luttinger liquid phase for $0<n<1$. On both figures, we indicate (arrows) the magnetic field 
at which the edge states are occupied by one magnon.}
\label{fig:mbnb}
\end{figure}

The magnetization curves from the spin model and the boson Hamiltonians are shown 
in Fig. \ref{fig:mbnb}(a). The models present two magnetization plateaus, one 
at the ferrimagnetic magnetization $m_{\scriptscriptstyle{FRI}}=S-1/2$ and 
the fully polarized state $m_{\scriptscriptstyle{FP}}=S+1/2$. 
The saturation field $B_{FP}$, end point of the FP plateau, is obtained 
through the closing of the single-particle magnon gap with $B$, as sketched 
in Fig. \ref{fig:magnonbands}, at $B=B_{\scriptscriptstyle{FP}}=2(S+s)$. 
The saturation field 
$B_{\scriptscriptstyle{FP}}$ from any bosonic approximation is rigorously equal to its 
exact value since the fully polarized state, an exact eigenstate of the spin model, 
is the vacuum for the bosonic models. 

{The Oshikawa-Yamanaka-Affleck topological criteria \cite{PhysRevLett.78.1984} states 
that a plateau can appear in the magnetization curve of spin systems if 
\begin{equation}
 S_u-m=integer,
 \label{oya}
\end{equation}
where $S_u$ is the maximum spin of a unit cell, unless the ground state spontaneously 
break translation symmetry. This corresponds to a number of magnons per unit cell $n=0,1,2,\ldots,S_u (S_u-1/2)$ for integer (half-integer) $S_u$, from the fully polarized state. 
Since $S_u=S+1/2$ (saturated magnetization) for spin-(1/2,$S$) chains, plateaus could 
appear at $m=S+1/2,~(S-1/2),\ldots,0(1/2)$, for 
half-integer (integer) $S$. In the spin-(1/2,$S$) chains, the 
data shows that there are two magnetization plateaus, one at the fully polarized ($n=0$) magnetization and the other at one magnon per unit cell ($n=1$), the ferrimagnetic plateau.   
The other possible magnetization plateaus are
inhibited by the magnon-magnon interaction term $V$ [Eq. (\ref{eq:hamiltonianhv})].
The magnetization curves of ferrimagnetic spin chains with $1/2<s<S$ can exhibit other plateaus between the ferrimagnetic and the fully polatized ones. 
This is observed, for example, in spin-(1,2) and spin-(1,3/2) chains \cite{Sakai2000}. For these chains, the ferrimagnetic plateau state has two magnons per unit cell, $n=2$, and the magnetization curves also exhibits a plateau at $n=1$. 
We also observe the occupancy of the edge states of 
the ferrimagnetic plateau by one magnon at the indicated magnetic fields. 
These edge states appears in finite-size open systems  
associated with topological aspects \cite{Montenegro-Filho2020, Hu2014,*Hu2015} of the ferrimagnetic state. 

The non-interacting Rice-Mele model (\ref{eq:ricemele}) does not present edge states for 
uniform hopping. 
However, recently \cite{Lin2020}, it was shown that an interacting \textit{fermionic system}, 
with a local Coulomb interaction, presents 
effective edge states, and that a fraction of the boundary charge is, in fact, related 
to the bulk properties. Our non-interacting Hamiltonian (\ref{eq:hamiltonianht}) 
has the modified local potentials in the boundaries (\ref{eq:localboundary}), required 
from the spin mapping, and thus localized edge states. Furthermore, our interacting model 
is a \textit{bosonic system} having the correlated hopping and nearest-neighbor coulomb repulsion. In the Sec. \ref{sec:rice-mele-edge}, we study the boundary magnon density and 
compare it from relevant interacting and non-interacting bosonic models.}

The gapless phase between the magnetization plateaus is a Luttinger liquid phase with power-law decay of the transverse spin correlation functions \cite{giamarchi2003quantum,*PhysRevB.55.5816} and has a dynamical exponent $z=1$.

We notice in Fig. \ref{fig:mbnb}(a) that as $B$ decreases from $B_{\scriptscriptstyle{FP}}$, 
the magnetization from the free hard-core model, $t$-approximation, 
departs from that of the spin model at roughly half filling of the 
lower magnon band $\omega^{\scriptscriptstyle{(-)}}$, Eq. (\ref{eq:dispersion}). At 
this filling, the interaction effects start to become relevant. 
The critical field of the ferrimagnetic plateau from the free 
hard-core model: $2S$, see Fig. \ref{fig:magnonbands}, 
becomes more near its value for the spin model as $S$ increases, as also confirmed 
in Table \ref{tab:fricriticalfield}.
This effect can be attributed to the local energy of the $a$ sites that 
becomes deeper in comparison with the local energy of the $b$ sites, as 
can be seen in the energy term (\ref{eq:hamiltonianht}) and the sketch 
in Fig. \ref{fig:models}. Thus, the magnons become  
more localized on $a$ sites as $S$ increases, and the $X$ and $V$ energy terms, 
Eqs. (\ref{eq:hamiltonianx}) and (\ref{eq:hamiltonianhv}), respectively, 
become lesser relevant.  The $X$ term, due to the low occupancy probability  
of one $b$ site by two magnons, and also because $X\rightarrow0$ as $S\rightarrow\infty$, 
see Eq. (\ref{eq:xterm}). 
The $V$ term, on the other hand, because 
the probability of finding two magnons in nearest neighbor
sites is also very low. Further, the $t-X-V$-approximation, which has 
all energy terms in Hamiltonian (\ref{eq:hamiltonianhp}) active, is in excellent 
agreement with the results for the spin model. Even the location of the edge states 
is well reproduced by the $t-X-V$-approximation. 

In Fig. \ref{fig:mbnb}(b) we show the average magnon density per unit cell $\langle n\rangle$.
Besides the models presented in (a), we add the $t-V$-approximation.
The presence of the magnon- repulsion makes the accordance 
with the spin model good for $\langle n\rangle > 1/2$, while in the $t$-approximation
this agreement is good up to a value of $\langle n\rangle$ less than 1/2.
As $\langle n\rangle\rightarrow 1$, reaching the ferrimagnetic plateau, the repulsion $V$ increases
much the energy of the system, and, thus, implies a lower value of the critical magnetic field
$B_{FRI}$, compared to the spin model. 
The location of the edge states in the finite-size system 
is also different between the spin and 
the $t-V$-approximation. Further, the $X$-term, the density-dependent hopping term, 
lowers the energy, and the full $\langle n\rangle$-curve of the spin model is
well reproduced by the $t-X-V$ approximation.  
{ We notice, however, that only for the finite size spin-(1/2, 1) chain studied, the lower critical field of the ferrimagnetic plateau is $B\approx 0.06J$. In Sec. \ref{sec:rice-mele-edge}, we present the magnon curve 
for $N>128$.}

\begin{table}
\caption{\label{tab:fricriticalfield} Critical field of the ferrimagnetic plateau 
in units of $J$ for the spin-(1/2, $S$) chains in the free hard-core boson 
approximation $t$-approximation and the spin model.}
\begin{ruledtabular}
\begin{tabular}{cccc}
\textbf{$S$} & $t$-approx.: 2$S$ & spin model: $\beta$& $\frac{2S-\beta}{\beta}$\\\hline
1 & 2 & 1.76 & 0.13\\
3/2 & 3 & 2.85 & 0.05\\
2 & 4 & 3.88 & 0.03\\
5/2 & 5 & 4.91 & 0.02\\
\end{tabular}
\end{ruledtabular}
\end{table}

\begin{table}
\caption{\label{tab:localmag} Average spins at $a$ ($\langle s^z\rangle$) and 
$b$ ($\langle S^z\rangle$) sites 
 for the spin-(1/2, $S$) chains in the free hard-core boson 
approximation $t$ and the spin model: 
$(\langle s^z\rangle,\langle S^z\rangle)$, at $m=S-s$, the ferrimagnetic magnetization.}
\begin{ruledtabular}
\begin{tabular}{ccc}
\textbf{$S$} & $t$-approx. & spin model \\\hline
1 & $(-0.27,0.77)$ & $(-0.29,0.79)$ \\
3/2 & $(-0.34,1.34)$ & $(-0.36,1.36)$ \\
2 & $(-0.38,1.88)$ & $(-0.39,1.89)$ \\
5/2 & $(-0.40,2.40)$ & $(-0.41,2.41)$ \\
\end{tabular}
\end{ruledtabular}
\end{table}

\begin{figure}
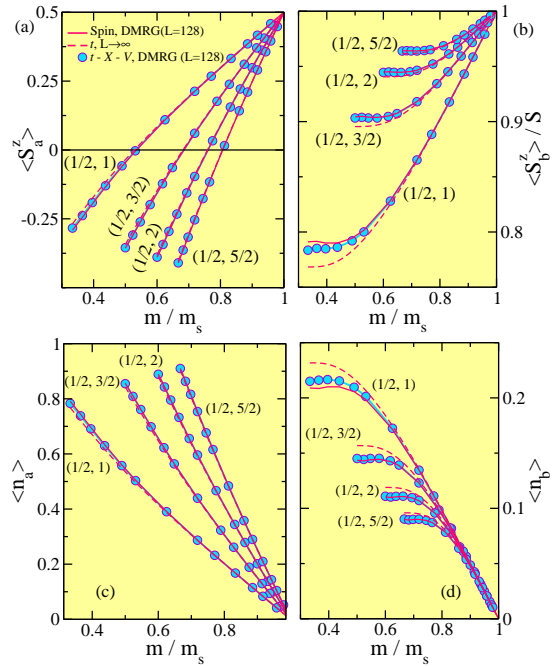

\includegraphics*[width=0.40\textwidth]{fig4ab.eps}
\includegraphics*[width=0.40\textwidth]{fig4cd.eps}
\caption{Average spin at (a) $a${, $\langle S^z_a\rangle$,} and (b) $b${, $\langle S^z_b\rangle$,} sites, in this 
case normalized by $S$, for (1/2, $S$) chains with 
$S=1,~3/2,~2,\text{ and }5/2$, as a function of the normalized 
magnetization $m/m_s$, where $m_s$ is the saturation magnetization of 
each chain. The hard-core boson $t$ approximation in the thermodynamic limit (dashed lines), 
DMRG results for the 
spin model  (full lines) and the $t-X-V$ approximation ($\color{Az75}\bullet$), both for
systems with $L=128$. Average magnon densities at 
(c) $a${, $\langle n_a\rangle$,} and (d) $b${, $\langle n_b\rangle$,} sites as a function of $m/m_s$ and 
the same legend of (a) and (b).}
\label{fig:localmag}
\end{figure}

The average spin and magnon density at $a$ and $b$ sites are shown
in Fig. \ref{fig:localmag}. Notice that the average spin at $b$ sites, Fig. \ref{fig:localmag}(b), 
is normalized by $S$. Also shown are the probability of occupancy of $a$ and $b$ sites. 
We use the expressions (\ref{eq:localmag}) to obtain the average spins from the $t$-approximation, 
while we calculate 
the average magnon density from the spin model with
\begin{equation}
 \begin{cases}
  \langle n_a\rangle=\frac{1}{2}-\frac{1}{L}\sum_{l=1}^{L}\langle s^z_l\rangle,\\
  \langle n_b\rangle=S-\frac{1}{L}\sum_{l=1}^{L}\langle S^z_l\rangle,
 \end{cases}
\end{equation}
for a finite spin chain of size $L$ and open boundaries.
The value of the average 
spin at $a$ sites from any of the considered bosonic approximations is in  
good agreement with its value from the spin model.  
A relevant departure between
the approximations and the spin model occurs in the average spin on $b$ sites as the ferrimagnetic
magnetization plateau is approached. 
However, even the $t$-approximation provides good values for 
the average spins, as shown in Table \ref{tab:localmag} for the ferrimagnetic magnetization.
Also, the results become indistinguishable, even on 
$b$ sites, as $S$ increases or as the fully polarized plateau is approached.
Furthermore, as in the magnetization results, the 
$t-X-V$ model is an excellent approximation to the spin Hamiltonian. 
The data in Figs. \ref{fig:localmag}(c) and (d) confirm that due to the higher value of the 
local potential on $a$ sites, the probability of occupancy of $a$ sites 
is higher than that on $b$ sites, as discussed in the context of the magnetization
curves in Fig. \ref{fig:mbnb}. In particular, quantum fluctuations are reduced as $S$
increases, since $\langle n_b\rangle\rightarrow 0$ in this limit, as 
the data in Fig. \ref{fig:localmag}(d) suggests.
\begin{figure*}[htb]
\includegraphics*[width=0.38\textwidth]{fig5a.eps}
\includegraphics*[width=0.38\textwidth]{fig5b.eps}
\includegraphics*[width=0.20\textwidth]{fig5c.eps}
\includegraphics*[width=0.8\textwidth]{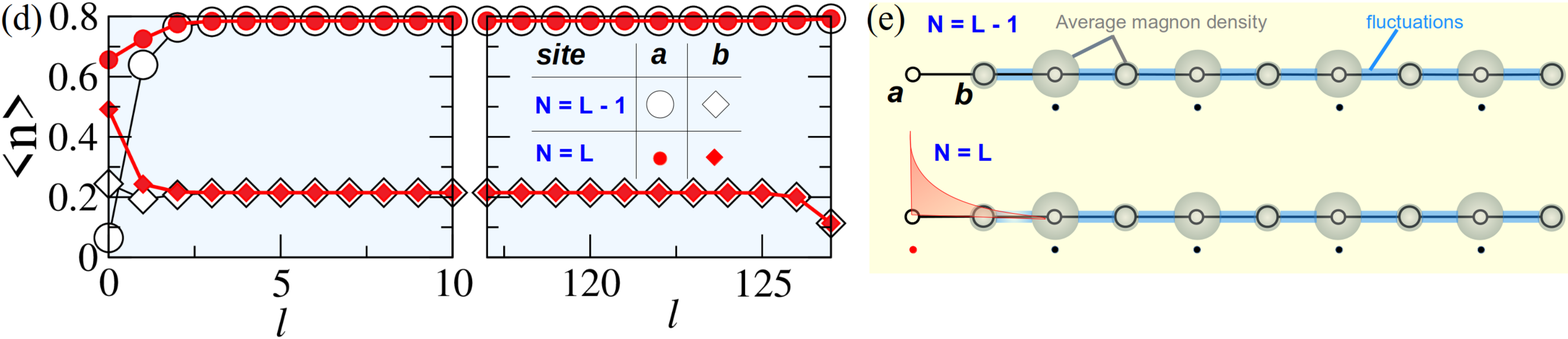}
\caption{Average magnon densities along (a) the ($1/2$, $S=1$) chain
and (b) the ($1/2$, $S=5/2$) with $L=128$ unit cells. Magnon densities at ($\color{Pnk102}\bullet$) $a$ sites, 
$\langle n_{a,l}\rangle$, at ($\color{Blu82}\bullet$) $b$ sites, $\langle n_{b,l}\rangle$,
and ($\color{Trq67}\bullet$) total magnon density, $\langle n_{cell,l}\rangle$, for unit cell $l$ from
the spin model and ($\blacktriangle$) the $t-X-V$ approximation, both calculated with DMRG for systems 
with $L=128$. The following 
total average magnon densities are shown: $n=8/128,~16/128,~72/128,
\text{ and }~116/128$, from top to bottom, corresponding, respectively,
to the magnetizations (a) $m=184/128,~176/128,~120/128,\text{ and }76/128$; and
(b) $m=376/128,~368/128,~312/128,\text{ and }268/128$. (c) Average magnon distribution change 
$\delta\langle n\rangle_{FRI+1\rightarrow FRI, l}=\langle n\rangle_{N=L,l}-\langle n\rangle_{N=L-1, l}$ 
along the 
chain as a function of unit cell $l$ shows the presence of the edge states for 
the four chains studied. (d) The average magnon probability density {$\langle n\rangle$} along a chain with 
$L=128$ at (circles) $a$ 
and (diamonds) $b$ sites for 
(white symbols) $N=L-1$ and (red symbols) $N=L$ bosons from the $t-X-V$ approximation of the 
(1/2,1) chain. 
(e) Sketch of the ground states for $N=L-1$ and $N=L$ bosons.
Average magnon densities are shown as grey circles, fluctuations between $a$ and $b$ sites are
indicated by a blue stripe, and we use a red filled curve to represent the localized orbital in
the boundary unit cell.} 
\label{fig:magnondens}
\end{figure*}

In Figs. \ref{fig:magnondens}(a) and (b) we present the local magnon densities in finite chains with 
open boundaries for the spin and the $t-X-V$ models. The magnon densities 
from the spin model are calculated through
\begin{equation}
 \begin{cases}
  \langle n^a_l\rangle = \frac{1}{2}-\langle s^z_l \rangle;\\
  \langle n^b_l\rangle = S-\langle S^z_l \rangle,\\
 \end{cases}
\end{equation}
and $\langle n^{cell}_l\rangle=\langle n^a_l\rangle+\langle n^b_l\rangle$. 
We show the data for the spin-(1/2, 1) and spin-(1/2,5/2) chains, for two 
low magnon densities ($n=8/128$, $n=16/128$), near the fully polarized magnetization plateau, 
and two high magnon densities, more near the 
ferrimagnetic magnetization plateau ($n=72/128$, $n=116/128$), in a chain with $L=128$
unit cells. 
For the lower magnon densities (hard-core limit) the results for $t-X-V$-approximation departs from the 
spin model data near the boundaries for the spin-(1/2, 1). However, the accordance between the two models
is excellent in the case of the spin-(1/2, 5/2), even near the boundaries, for the two
lower magnon densities shown. For the two higher magnon densities, there is an excellent agreement 
between the spin model and the $t-X-V$ approximation for the two chains. 

In Fig. \ref{fig:magnondens}(c), we show the magnon distribution in the edge localized state
occupied by one magnon for the four chains studied. To calculate it, we notice that the edge state appears
between the two magnetization steps: $S^z=L(S-s)$ and $S^z=L(S-s)+1$, which will join in the thermodynamic limit,
and make the $S-s$ ferrimagnetic plateau, see Fig. \ref{fig:mbnb}(a) and (b).
Thus, to visualize the spatial extent of the edge state, we consider the magnon distribution change, $\delta\langle n\rangle_{FRI+1\rightarrow FRI}$, between a total number of magnons $N=L-1$ [$S^z=L(S-s)+1$] and 
$N=L$ [$S^z=L(S-s)$]:
\begin{equation}
\delta\langle n\rangle_{FRI+1\rightarrow FRI, l}=\langle n\rangle_{N=L,l}-\langle n\rangle_{N=L-1, l}.
\end{equation}
We notice in the data shown in Fig. \ref{fig:magnondens}(c) that a tiny discrepancy is 
observed between the results for the $t-X-V$-approximation and the spin model 
in the case of the (1/2, 1) chain, while for the other chains the agreement is excellent.
The hole added to the many-magnon state at $N=L$ becomes well localized in the 
boundary cell, mainly on the boundary $a$ site, thus implying the presence of this empty 
orbital in the $N=L-1$ [$S^z=L(S-s)$] state. The localization in the boundary increases 
with $S$ as expected from the increasing of the bulk gap.
{ In fact, the fluctuations of the magnon 
density in the boundary become neglegible as $S\rightarrow \infty$, and  
the boundary hosts one magnon in this limit. 
We can attibute this behavior to the increasing potential barrier between bulk and edge, Eqs. (\ref{eq:httransform}) and (\ref{eq:localboundary}), respectively, as $S\rightarrow\infty$.} 

As an example of the data used to make the results shown in Fig. \ref{fig:magnondens}(d), 
we exhibit in Fig. \ref{fig:magnondens}(d) 
the average magnon density along the chain, $\langle n\rangle$, for $a$ and $b$ sites, in the 
$t-X-V$ approximation of the (1/2,1) chain. We notice that the $a$ site at the left boundary
is almost empty for $N=L-1$, while its $\approx 0.6$ for $N=L$, implying the value shown in (c).
We also notice a sizable change in the occupation of the $b$ site at the left boundary for the two fillings,
while $\langle n\rangle$ does not change at the right boundary. 
A sketch of the two ground states is shown in 
Fig. \ref{fig:magnondens}(e). For $N=L-1$, we have an insulating state with the magnon average 
higher on $a$ sites than on $b$ sites. Quantum fluctuations between $a$ and $b$ sites in 
nearby unit cells are expected, 
given the excellent accordance between the results from the $t-X-V$-approximation and the spin model
in the high-density limit, $m\rightarrow m_{FRI}$. Adding one magnon to the $(L-1)$-magnon state, we obtain the magnetization 
at $m=S-s$ in the finite-size system, which is also an insulating state. The added magnon occupies 
the empty localized orbital state in the unit cell of the boundary $a$ site. The penetration of this
edge state in the gapped bulk is very tiny and decreases with increasing $S$, see Fig. \ref{fig:magnondens}(c), 
as can be estimated through the average magnon distribution.

{
\section{The relevance of the boundary potentials and interactions in the 
boundary magnon density of the bosonic model}
\label{sec:rice-mele-edge}
With the help of Fig. \ref{fig:ricemele-mod}, we discuss 
the relevance of the difference between the local potential in the boundaries 
and the bulk sites on the boundary magnon densities, as well as
the importance of interactions in it. 
We use the parameters of the spin-(1/2, 1) chain, but 
focus on the bosonic Hamiltonian, for three models in finite-size chains: 
(1) the interacting model considering the mapping to the spin system ($X\neq 0, V\neq 0$) - the $t-X-V$ model; (2) the corresponding non-interacting model, i. e., the 
$t-X-V$ model with $X=0$ and $V=0$, which has a hard-core constraint on $a$ sites and 
the $b$ sites can host up to 2 bosons; and (3), the $t$ approximation, which has a 
hard-core constraint on $a$ and $b$ sites. One of the sets, Figs. \ref{fig:ricemele-mod}(a)
and (c), shows data for the boundary potentials distinct from the bulk potentials, with 
the values in Eqs. (\ref{eq:localboundary}) and (\ref{eq:httransform}), respectively.
The other set, Figs. \ref{fig:ricemele-mod}(b) and (d), takes the value of the 
boundary potentials equal to the bulk ones, Eq. (\ref{eq:httransform}).  
The arrows in Figs. \ref{fig:ricemele-mod}(a) and (b) indicate the edge states whose 
boson density change
\begin{equation}
\delta\langle n\rangle_{N_i\rightarrow N_f,l}=\langle n\rangle_{N_f,l}-\langle n\rangle_{N_i, l}
\end{equation}
is shown in Figs. \ref{fig:ricemele-mod}(c) and (d), respectively, where $N_f$ ($N_i$) is 
the highest (lowest) value of $N$ between the two magnon density steps separated by
the occupancy of the edge state.

\begin{figure}
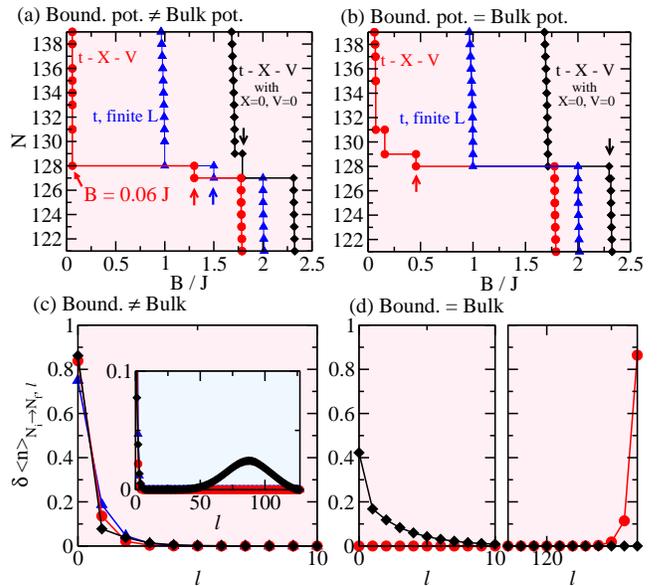

\includegraphics*[width=0.47\textwidth]{fig6ab.eps}
\includegraphics*[width=0.47\textwidth]{fig6cd.eps}
\caption{{ DMRG results for finite-size modified Rice-Mele models 
with $L=128$. In (a) and (c), the boundary potentials have the values in 
Eq. (\ref{eq:localboundary}); 
while in (b) and (d), the boundary potentials have the same values of the bulk local potentials on $a$ and $b$ sites, 
$\varepsilon_a$ and $\varepsilon_b$ in Eq. (\ref{eq:httransform}). 
Taking the parameters from the 
spin-(1/2, 1) chain, we show data for the full interacting model, $t-X-V$, 
the $t-X-V$ model with $X=0=V$, and the hard-core approximation $t$. 
In (a) and (b), we present the number of magnons $N$ as a function of 
magnetic field (in units of $J$) $B=-\mu$, where $\mu$ is the chemical potential in 
the bosonic system. In (c) and (d), we show the change in magnon 
density per unit cell $\delta\langle n\rangle_{N_i\rightarrow N_f,l}=\langle n\rangle_{N_f,l}-\langle n\rangle_{N_i, l}$ 
between states $N=N_i$ and $N=N_f$ as a function of unit cell $l$, where 
the arrows in (a) and (b) identify $N_i$ and $N_f$ for each model.}}
\label{fig:ricemele-mod}
\end{figure}
For the bulk different from the boundaries, Fig. \ref{fig:ricemele-mod}(a), we see that the edge state
is very robust, appearing even in the $t$ approximation. This does not occur for the $t$ approximation if the 
boundary is equal to the bulk, Fig. \ref{fig:ricemele-mod}(b), which 
is an expected result for the Rice-Mele model \cite{Lin2020}. However, we notice in Fig. \ref{fig:ricemele-mod}(a) that, in comparison with the spin model, 
the plateau width is very underestimated in the $t$ approximation, since we
are not discarding the single-particle states in the upper band shown in Fig. \ref{fig:magnonbands}. We observe that this
upper band was discarded for the $L\rightarrow\infty$ results shown in Figs. \ref{fig:mbnb} and \ref{fig:localmag},
due to the introduction of the effective chemical potential connected with the Takhashi's constraint \cite{DaSilva2017}. Considering also the spin model, the
error in the plateau width for $t-X-V$ interacting case is very tiny, with a lower
critical field $\approx 0.06 J$. Also in Fig. \ref{fig:ricemele-mod}(a), bulk and boundaries distinct, we note that the edge state in the non-interacting $t-X-V$ model appears 
only for two bosons over the $N=127$ state, Fig. \ref{fig:ricemele-mod}(a), with one of the bosons occupying an extended
state and the other an edge state, as shown in Fig. \ref{fig:ricemele-mod}(c). This shows that the distinction between
boundaries and bulk 
is not sufficient to a good representation of the edge state in the spin model, the \textit{interactions} are essential for it. 
In the case of the $t$ approximation, the particle statistics provides a change in the single particle
quantum states that mimic the effect of the $X$ and $V$ interactions in the $t-X-V$ model. In Fig. \ref{fig:ricemele-mod}
(b), where the boundary is equal to the bulk, we show that 
edge states appear in the $t-X-V$ model in the interacting
and non-interacting cases. As mentioned above, the edge state of the non-interacting case 
does not appear in the $t$ approximation.  
Further, the boundary density, Fig. \ref{fig:ricemele-mod}
(d), in the interacting $t-X-V$ case appears in the right
extreme and has a value approximately equal to that in Fig. \ref{fig:ricemele-mod}
(c). From these results, we can assert that the interactions $X$ and $V$, and the particle statistics, 
are essential to understand the boundary magnon density.}

\section{Summary and discussion}
We have investigated the Holstein-Primakoff bosonic Hamiltonian, up to order $\sqrt{s/S}$, of 
the alternating spin-($s=1/2$, $S$) chains in a magnetic field, and considering the fully polarized
as the vacuum. Three bosonic Hamiltonians were considered: the first 
has only a hopping term ($t$-approximation) and distinct local potentials ($\varepsilon$) on the spin-$1/2$ ($a$ site)
and the spin-$S$ ($b$ site) sites {, this approximation is a Rice-Mele Hamiltonian with boundaries different from the bulk and uniform hopping term}. The second one has the terms of the $t$-approximation plus a magnon-magnon repulsion $V$ ($t$-$V$-approximation); and the last approximation shows the $t$ and $V$ terms, and a density-dependent correlated
hopping term $X$ ($t-X-V$-approximation). 
In the $t-X-V$ approximation, $b$ sites can accommodate 
up to two magnons, while in the others a hard-core constraint is imposed on $a$ and $b$ sites. 
The local potentials, at any density, and $V$ for higher densities favor 
magnon localization on the $a$ sites. On the other hand, quantum fluctuations, 
magnon tunneling between $a$ and $b$ sites, are favored by 
$t$, at any density, and $X$ for higher densities.
We use the density matrix renormalization group
to investigate the spin model and the bosonic Hamiltonians $t-V$ and $t-X-V$ in
finite-size open systems, while for the $t$-approximation we have considered its analytical solution. 
We compare the magnetization and magnon densities per unit cell
as a function of a magnetic field, average bulk density and local densities along the chains from the spin model and the bosonic approximations. 
From the ferrimagnetic plateau (one magnon per unit cell) to saturation (empty chain), 
the $t-X-V$-approximation is in excellent agreement with the spin model, 
while the $t$ and $t-V$ results depart from that of the spin model as the magnon density increases. 
This, thus, shows the relevance of both interaction terms, magnon-magnon repulsion and 
the correlated hopping term $X$, and its associated particle fluctuations, 
to describe the many-body system near the ferrimagnetic plateau.
{ The edge state associated with the insulating ferrimagnetic plateau 
is well reproduced by the bosonic model. In particular, we have shown that the 
magnon boundary densities are strongly dependent on the interactions and the particle statistics.}    

{The use of the fully polarized state as the vacuum enables a better understanding of the underlying quantum processes 
in the spin Hamiltonian, compared to a ferrimagnetic vacuum, which is not an exact eigenstate of the spin Hamiltonian. Our results also suggest that, beyond hopping and magnon-magnon interaction, the density-dependent hopping term increases the range of magnetizations for which effective bosonic models can make a good description of the physical data from general spin systems having ions with spins higher than 1/2.}

\begin{acknowledgments}
We acknowledge support from Coordena\c{c}\~ao de Aperfei\c{c}oamento de Pessoal de N\'{\i}vel Superior (CAPES), Conselho Nacional de Desenvolvimento Cient\'{\i}fico e Tecnol\'ogico (CNPq), and Funda\c{c}\~ao de Amparo \`a Ci\^encia e Tecnologia do Estado de Pernambuco (FACEPE), Brazilian agencies, including the PRONEX Program which is funded by CNPq and FACEPE, APQ-0602-1.05/14. 
\end{acknowledgments}

\end{document}